\documentclass[structabstract]{aa}

\usepackage{graphicx}
\usepackage{txfonts}

\usepackage{longtable}
\usepackage{natbib}
\newcommand{\kms}{km\,s$^{-1}$}

\newcommand{\uchii}{{UC H{\scriptsize II}}}
\newcommand{\hii}{{H{\scriptsize II}}}
\newcommand{\mjybeam}{$\mbox{mJy beam}^{-1}$}

\newcommand{\Lsun}{\mbox{\,$\rm L_{\odot}$}}
\newcommand{\Msun}{\mbox{\,$\rm M_{\odot}$}}

\usepackage{cite}
\begin{document}
   \title{Dynamics of the 6.7 and 12.2~GHz  methanol masers around Cepheus~A HW2}

   \subtitle{}

   \author{Karl J.E. Torstensson\inst{1,2}
	\and
	Huib Jan van Langevelde\inst{2,1}
        \and
	Wouter H.T. Vlemmings\inst{3}
	\and 
	Stephen Bourke\inst{2}
        }


   \institute{Leiden Observatory, Leiden University,
              PO Box 9513, 2300 RA Leiden, The Netherlands\\
              \email{kalle@strw.leidenuniv.nl}
         \and
             Joint Institute for VLBI in Europe,
             PO Box 2, 7990 AA Dwingeloo, The Netherlands\\
             \email{[langevelde;bourke]@jive.nl}
        \and
	    	Argelander-Institut f\"ur Astronomie, University of Bonn\\
	     	Auf dem H\"ugel 71, D-53121 Bonn, Germany
	     	\email{wouter@astro.uni-bonn.de}
	 }

   \date{Received August 13, 2010; accepted October 15, 2010}


  \abstract
   {The 6.7 GHz methanol maser is exclusively associated with high-mass star formation. However, it remains unclear what structures harbour the methanol masers. Cepheus A is one of the closest regions of massive star formation, making it an excellent candidate for detailed studies.}
   {We determine the dynamics of maser spots in the high-mass star-forming region Cepheus A in order to infer where and when the maser emission occurs.}
   {Very long baseline interferometry (VLBI) observations of the 6.7 and 12.2~GHz methanol masers allows for mapping their spatial and velocity distribution. Phase-referencing is used to determine the astrometric positions of the maser emission, and multi-epoch observations can reveal 3D motions.}
   {The 6.7 GHz methanol masers are found in a filamentary structure over $\sim$1350~AU, straddling the waist of the radio jet HW2. The positions agree well with previous observations of both the 6.7 and 12.2~GHz methanol masers. The velocity field of the maser spots does not show any sign of rotation, but is instead consistent with an infall signature. The 12.2~GHz methanol masers are closely associated with the 6.7~GHz methanol masers, and the parallax that we derive confirms previous measurements.}
   {We show that the methanol maser emission very likely arises in a shock interface in the equatorial region of Cepheus A HW2 and presents a model in which the maser emission occurs between the infalling gas and the accretion disk/process.}

   \keywords{Stars: formation - Masers - ISM: individual: \object{Cepheus A} - ISM}

   \titlerunning{Dynamics of the methanol masers around Cep~A HW2}
   \maketitle
%

\section{Introduction}
Methanol masers are signposts of high-mass star formation. Continuum emission of warm dust at sub-millimetre wavelengths has been detected at well over 95\% of the observed 6.7~GHz methanol maser sites \citep{hill05}, even though only some of the masers are associated with detectable ultra-compact (UC) \hii\ regions. This seems to indicate that the methanol masers probe a range of early phases of massive star formation and that they disappear as the \uchii\ region evolves \citep{walsh98}. More than just signposts, masers are tracers of the geometry and small-scale dynamics in these regions. Much effort has focused on studying the kinematics, and several claims of circumstellar disks, expanding spherical shells, and jets have been made \citep[e.g.][]{norris98a, pestalozzi04, minier02, bartkiewicz05b}. The evidence suggests that masers trace an evolutionary sequence or possibly a mass range in the formation process, or perhaps even both \citep{walsh01a}.

The star-forming region Cepheus A East (hereafter Cep~A) is one of the closest high-mass star-forming regions at a distance of 700~pc, as determined by parallax measurements of methanol masers \citep[hereafter Mos09]{moscadelli09}. The region has a total bolometric luminosity of $2.5 \times 10^4$\Lsun\ \citep{evans81}, and its appearance in the radio is dominated by the thermal jet Cep~A HW2 \citep{hw84}. A central object has been identified that is believed to be the source driving the jet. On small scales ($\sim1$\arcsec), the thermal jet shows outflow velocities in excess of 500 \kms\ \citep{curiel06}. Also on larger scales ($\sim$1\arcmin), a bipolar molecular outflow with blueshifted gas in the NE and redshifted gas in the SW is seen in HCO$^+$ \citep{gomez99a}. Recent sub-millimetre observations have shown an elongated disk structure \citep{patel05, torrelles07} at the position of the 7~mm continuum object identified by \citet{curiel06}. Although there are multiple sources within the inner 1\arcsec, the central object (HW2) with a mass of $\sim$18~\Msun\ is believed to be the main driving source in the region \citep{jimenez-serra09a}.

Maser observations of HW2 have shown a complex structure in hydroxyl \citep{bartkiewicz05a}, water \citep{vlemmings06}, and 12~GHz methanol masers \citep[hereafter Min01]{minier01}. Recent multi-epoch observations of water masers in the region indicate the presence of a slower wide-angle outflow in addition to the high-velocity collimated jet observed in radio continuum \citep{torrelles10a}. \citet{sugiyama08a} find the 6.7~GHz methanol maser emission to be constrained to two individual clumps with blueshifted masers to the east and redshifted masers in a linear configuration to the west. Moreover, single-dish monitoring of the 6.7~GHz methanol masers has shown the variation in the maser emission in the blueshifted and redshifted cluster to be synchronised and anti-correlated \citep{sugiyama08b}. Maser polarisation observations indicate a strong ($|B| \sim 23$~mG) magnetic field in the 6.7~GHz methanol maser region that is aligned with the thermal jet of the protostar HW2 \citep{vlemmings08a, vlemmings10a}. Similarly to the molecular and dust disk, the methanol masers are aligned perpendicular to the thermal jet and magnetic field although at a larger radius. The inferred high accretion rate \citep{mckee03a} is likely regulated and sustained by the large-scale magnetic field.

We carried out large field of view European VLBI Network\footnote{The European VLBI Network is a joint facility of European, Chinese, South African and other radio astronomy institutes funded by their national research councils.} (EVN) observations of the 6.7~GHz and Very Long Baseline Array (VLBA), operated by the National Radio Astronomy Observatory \footnote{The National Radio Astronomy Observatory is a facility of the National Science Foundation operated under cooperative agreement by Associated Universities, Inc.} (NRAO), observations of the 12.2~GHz methanol masers in Cep~A East to determine the morphology and relationship between the 6.7~GHz and 12.2~GHz methanol masers. In Sec. \ref{observations} the observations and data reduction process is described. We report our results in Sec. \ref{results}, and extend the analysis to a simple geometric model in Sec. \ref{analysis}. The implications of our model and how it can be tested are discussed in Sec. \ref{discussion}, and finally our conclusions are presented in Sec. \ref{conclusions}.

\section{Observations and Data Reduction}\label{observations}
\subsection{6.7~GHz Data}

\begin{table}[h]
\caption{\label{obstable} Coordinates of science target and calibrators for the 6.7~GHz (EVN) and 12.2~GHz (VLBA) observations.}
\centering
\begin{tabular}{lll}
\hline\hline
Source			& RA(J2000)		& Dec(J2000)\\
\hline
Cep~A (HW2)	 	& 22 56 17.9000   	& +62 01 50.000 \\
J2302+6405		& 23 02 41.315001	& +64 05 52.84853 \\
3C345			& 16 42 58.809968 & +39 48 36.99400\\
DA193			& 05 55 30.805614	& +39 48 49.1650\\
\hline
Cep~A (HW2)		& 22 56 18.104	& +62 01 49.419  \\
J2232+6249		& 22 32 22.8655	& +62 49 36.436 \\
J2302+6405		& 23 02 41.315001	& +64 05 52.84853 \\
J2202+4216		& 22 02 43.291372	& +42 16 39.97992 \\
\hline
\end{tabular}
\tablefoot{ The top panel shows the calibrators and coordinates used for the 6.7~GHz observations conducted on Nov 6, 2004 with the EVN. The bottom panel displays the calibrators and coordinates of the 12.2~GHz observations carried out on 2006 March 3, June 2 and September 13, and 2007 January 28 with the VLBA.\\
}
\end{table}

The 6.7~GHz methanol maser observations of Cep~A were carried out with the EVN in November 2004 as a part of a larger project (EL032) with 12 sources in total. The eight telescopes of the EVN participating in the experiment were Medicina, Onsala, Toru\'n, Cambridge, Darnhall, Noto, Effelsberg and Westerbork. In order to achieve astrometric positions all the observations were done in phase referenced mode. For Cep A the calibrator J2302+6405 (with a separation of $\sim2.2$\degr) was used as the phase reference source. Cep~A was observed for a total of $\sim 2$~h including the phase calibrator and four scans on the amplitude and bandpass calibrators 3C345 and DA193. In Table \ref{obstable} we list the coordinates used for the calibrators and the phase centre of the observations. In order to improve the uv-coverage the observation was split into two $\sim$ 1h blocks. Unfortunately, due to scheduling constraints, these two blocks are for Cep~A separated by $\sim$ 12h, resulting in a less than ideal uv-coverage. The experiment was set up at a rest frequency of 6668.5142~MHz, with 1024 channels and with RCP and LCP recorded separately. The total bandwidth was 2MHz, resulting in a velocity resolution of 0.088~\kms, and a total velocity coverage of 90~\kms\ centred at LSR -3~\kms.

The data was correlated on the EVN correlator at JIVE with an integration time of 0.25~s. The short integration time was chosen to maximise the field of view (FOV) of the observations to enable a search for maser emission in a large field.

Several telescopes show internally generated RFI in the auto-correlated data. The strongest of these components exhibit Gibb's ringing, indicating that it is a very narrow-band signal. None of this internally generated RFI does however show in the cross-correlated data and is therefore not cause for any concern in the cross-correlation calibration.

All editing, calibration and imaging was done in AIPS\footnote{Astronomical Image Processing System, developed and maintained by the NRAO.}. The calibrators 3C345 and DA193, observed at the beginning and end of each 1~h block, were used for amplitude and bandpass calibration of each respective block. RCP and LCP data was edited and calibrated separately before combining them in the final image cube. During each 1~h block, eight cycles of Cep A and J2302+6405 were observed with an integration time of 3 and 2~min, respectively. After initial self calibration on J2302+6405 the phase solutions were transferred to the Cep~A data. The brightest maser feature was subsequently imaged to determine its absolute position after which it was used for further self calibration. After the final calibration had been applied a cleaned image cube was created for the central 64 channels for which maser emission could be seen in the channel spectra. The final images with a size of 2\arcsec $\times$2\arcsec\ and a pixel size 1~mas have an rms of 7~\mjybeam\ in the line-free channels and 60~\mjybeam\ in the channels with the brightest maser emission. 

In our analysis we are mostly interested in the relative astrometry with respect to the nearby (2.2\degr) phase calibrator. The accuracy of this will be limited mostly by the signal-to-noise of the maser and the atmospheric conditions, for which we have not done any specific calibration. Based on the analysis by \citet{rygl10a} we estimate the accuracy to be 0.3~mas in right ascension and somewhat worse in declination. The absolute astrometry has an additional component from the accuracy of the calibrator position and may be slightly larger than 1 mas. The amplitude calibration we estimate to be accurate to $\sim$30\% and the final image has a dynamic range of $\sim$1200 in the channel with the brightest emission. 

Additionally, we searched a much larger area for new methanol masers. The large field search was implemented using ParstelTongue \citep{kettenis06a} in which the field was split up in facets \citep{bourke06a}. Each of the 4669 boxes corresponds to a data cube of dimensions 2048$\times$2048 cells with a pixel size of 1~mas and 1024 frequency channels giving a total image size of over 18 terapixels ($\sim$2.5~arcmin diameter). This covers the half power beam width (HPBW) of baselines consisting of Effelsberg and one of the 32m dishes. The correlation parameters allow for imaging an area three times the beam size but sensitivity drops substantially as Effelsberg's contribution reduces. 

\subsection{12.2~GHz Data}
The multi-epoch VLBA observations of the 12.2~GHz methanol masers were performed on the dates as listed in Table \ref{obstable} (project BV059 A, B, C, D). All ten antennas of the VLBA participated in all epochs of the experiment. Each epoch consisted of a total of 5~h observing time. The receivers were setup to record two 0.5~MHz bands with 128 channels each, centred on the rest frequency of the maser line (12178.595~MHz), and one 4~MHz band on either side for the wide-band calibration. For each epoch the observing time was split up in two blocks in which during the first $\sim$1.5~h fast switching was done between Cep~A and two phase reference sources to ensure accurate astrometric positions. The two sources used for phase referencing were J2232+6249 (2.9\degr\ East) and J2302+6405 (2.2\degr\ North) of Cep~A and the switching sequence used was J2232+6249 - Cep~A - J2302+6405 with 40~s spent on each source. During the second block Cep~A was observed in four 31~min ``stares'', with each stare preceded by a 9~min observation of the bright calibrator J2202+4216 (BL Lac). The channel separation of 3.91~kHz for the narrow band data results in a velocity resolution of 0.96~\kms. Both RCP and LCP were recorded. The data were correlated at the VLBA correlator in Socorro with an integration time of 2~s. 

The initial calibration of amplitude, bandpass and delay was done on \object{J2202+4216} using standard AIPS procedures. The phase calibrators \object{J2232+6249} and \object{J2302+6405} were not bright enough to image directly and therefore reverse phase referencing was done on the brightest maser channel. Because the data was taken in a mixed bandwidth setup and full polarisation, a complex phase transfer scheme was implemented in ParselTongue \citep{kettenis06a} that extrapolates the phase and phase rate solutions from the maser channels to the bracketing continuum bands. This way both calibrators could be mapped in all epochs. The offset of the two phase reference sources was determined and applied to the Cep~A results. Average offsets were $\Delta\alpha=34.48$~mas and $\Delta\delta=67.87$~mas for J2232+6249, and $\Delta\alpha=49.52$~mas and $\Delta\delta=23.25$~mas for J2302+6405. The brightest maser features were fitted with a beam size of 3.5$\times$1.5~mas. Typical jmfit errors are 0.15~mas in right ascension and 0.1~mas in declination. However, the maser sources are quite elongated in right ascension (6~mas half power beam width vs 2.5 in declination) and taking into account the signal-to-noise of the calibrators and phase transfer effects we estimate the positional uncertainty relative to the calibrator to be $\sim$1~mas in right ascension and $\sim$0.2~mas in declination.

\section{Results}\label{results}
\subsection{6.7~GHz results}
\begin{figure*}
  \centering
  \includegraphics[angle=-90, width=\textwidth]{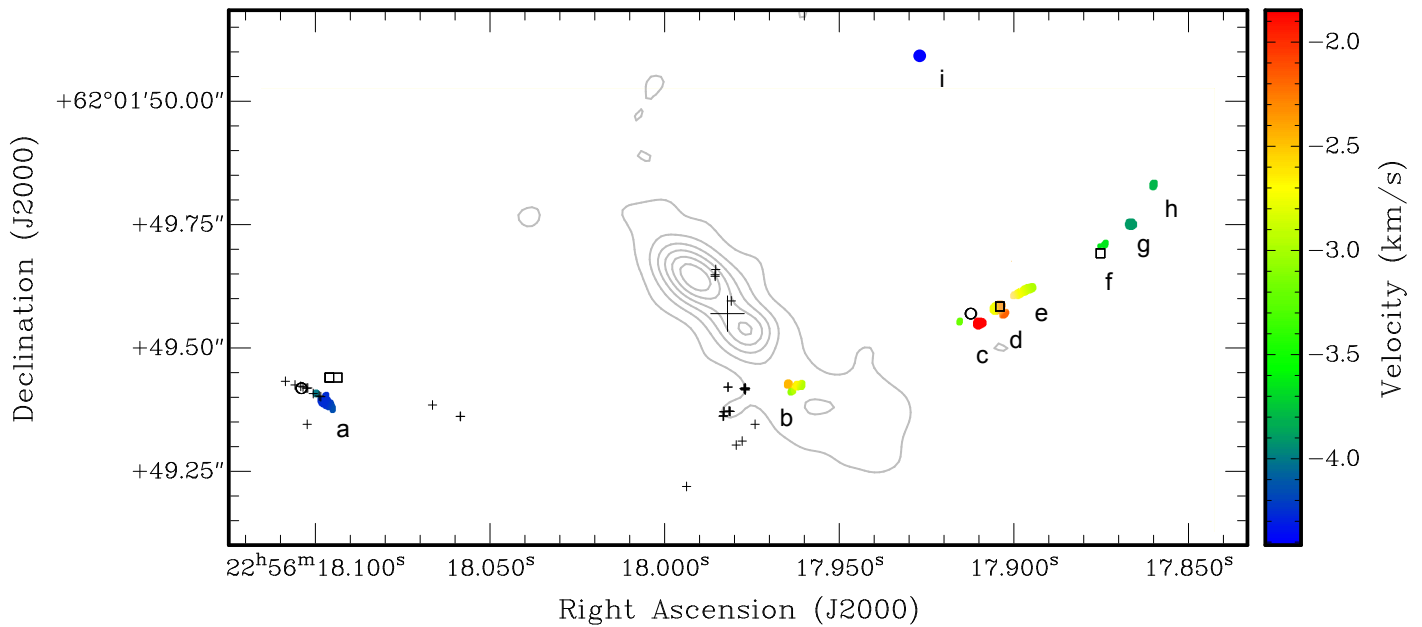}
  \caption{The velocity field of the 6.7~GHz methanol masers as obtained from a first moment map of the bright methanol emission (colour) with labels \emph{a}-\emph{i} overlaid on K band continuum (contours) \citep{torrelles98}. The point indicating the weak feature \emph{i} has been added to the moment map manually based on the results of detailed fitting of each channel. Also shown are the 12.2~GHz methanol masers (circles) in this paper and (boxes) Min01, and 22~GHz water masers (small pluses) \citep{vlemmings06}. The position of the protostar is indicated by the large plus sign \citep{curiel06}.}
  \label{cepamap}
\end{figure*}

The 6.7~GHz methanol maser emission in Cep~A originates in an extended $\sim$1.9$''$ (1350~AU) filamentary, arc-like structure, straddling the waist of the radio-jet HW2 (Fig. \ref{cepamap}). No other emission regions were found in the larger search.

The maser emission arises in nine elongated maser clumps separated spatially and/or in velocity. Their coordinates and LSR velocities are presented in Table \ref{masertable}. The individual maser features are elongated in the direction of the overall structure, strongly suggesting that the methanol maser in Cep~A is originating in one large scale structure. 

\begin{figure}
\centering
\includegraphics[width=9cm]{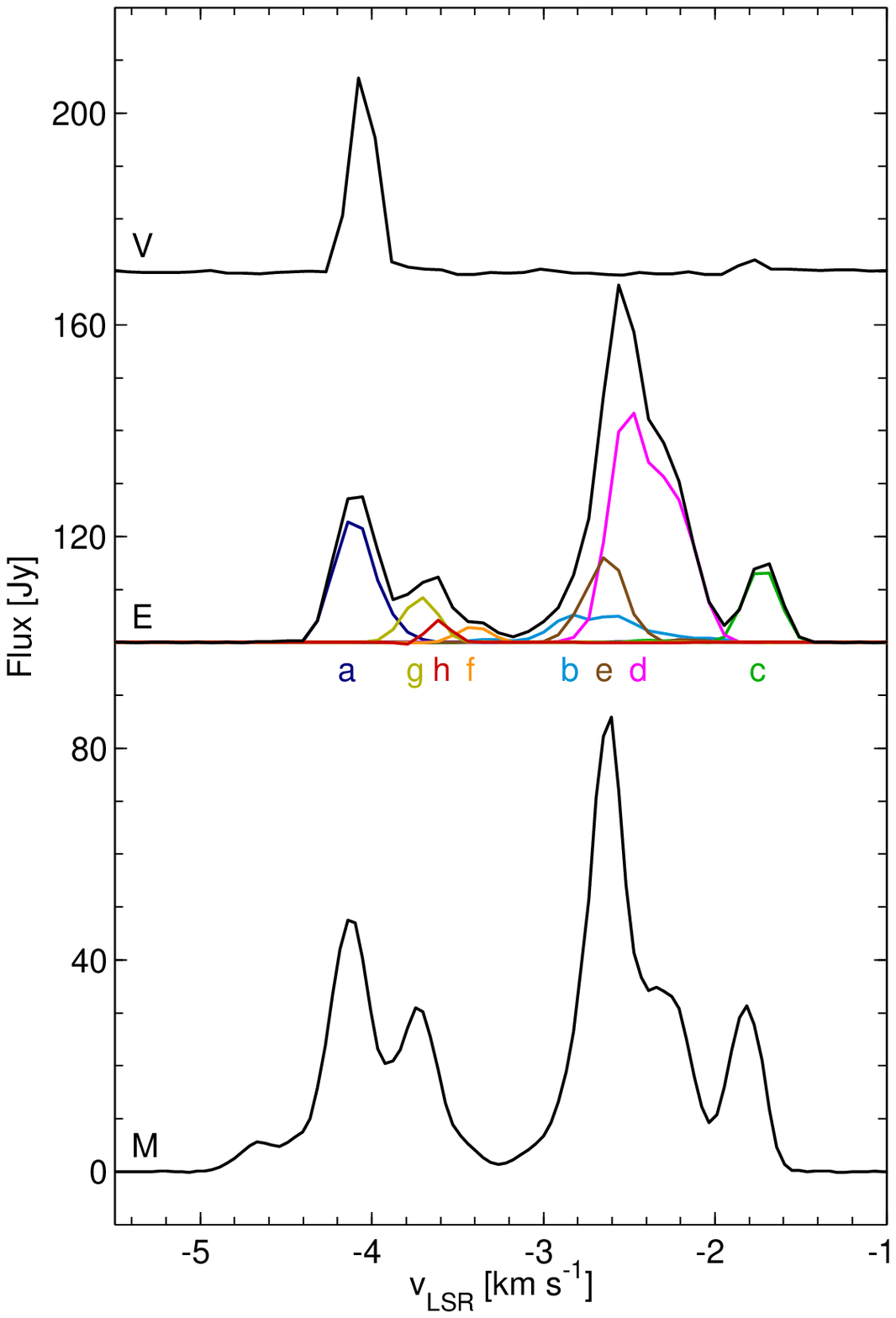}
\caption{\emph{Bottom}: (M) Merlin 6.7~GHz maser spectrum scaled down by a factor of 5 \citep{vlemmings10a}. \emph{Middle}: (E) Our 6.7~GHz total maser spectrum (black line) and individual spectra of maser clumps, from East (a) to West (h) as measured by the EVN. \emph{Top}: (V) A representative 12.2~GHz maser spectra measured with the VLBA scaled up by factor of 5.}
\label{maserspectra}
\end{figure}

There is a small velocity gradient of $\sim2.5 \times 10^{-3}$~\kms\,AU$^{-1}$ across the whole structure with higher velocities towards the centre. One maser feature (resolved into a few small spots) appears projected against the thermal jet seen in the radio continuum, see Fig. \ref{cepamap}. Its position is not coincident with the 7~mm continuum object identified by \citet{curiel06}, who claim this to be the driving source in HW2. Rather, the maser feature is seen further to the SW along the axis of the jet. In Fig. \ref{maserspectra} we show the total measured 6.7~GHz methanol maser spectrum and the measured MERLIN spectrum scaled down by a factor of five \citep{vlemmings10a}, also indicated are the individual spectra of the eight identified maser features. The emission is constrained to the velocity interval $-1.5$~\kms\ and $-4.5$~\kms. The morphology and velocity field that we find of the 6.7~GHz methanol masers agree well with previous measurements using different telescopes: \citet{vlemmings10a} (MERLIN); \citet{sugiyama07a,sugiyama08a, sugiyama08b} (VERA). One notable difference between our spectrum and earlier measurements is the lack of a feature at $\sim$-4.6~\kms\ as seen by both \citet{vlemmings10a} and \citet{sugiyama08b}. The missing feature corresponds to a clump of masers to the NNE of the brightest maser feature (\emph{d}). The missing maser spots are not bright enough to be seen in the spectrum or directly in the moment map. However, when imaging the cube channel by channel we were able to detect the faint emission and have included the feature \emph{i} in Fig. \ref{cepamap}. In Fig. \ref{ringmodel} these spots are more clearly visible.

\begin{table}[h]
\caption{The absolute positions and LSR velocities of the centroids of the brightest 6.7~GHz methanol maser features measured using EVN.}
\label{masertable}
\centering
\begin{tabular}{lcccccc}
\hline \hline
ID& RA &  Dec & v$_\mathrm{LSR}$ & Min01 \\
~ &(J2000) & (J2000) & [km\,s$^{-1}$] & designation \\
~ 	& 22:56: 		& 62:01:   	& ~ 		&~\\
\hline
a 	& 18.09648	& 49.4049	& -4.14	& A \\	
b 	& 17.96204	& 49.4369	& -2.56	& ~ \\	
c 	& 17.91032	& 49.5553	& -1.68	& ~ \\	
d 	& 17.90515	& 49.5843 	& -2.47	& B \\	
e 	& 17.89870 	& 49.6124	& -2.56	& ~ \\	
f 	& 17.87421	& 49.7063	& -3.44	& C \\	
g 	& 17.86687 	& 49.7442	& -3.70	& ~ \\	
h 	& 17.86038	& 49.8191	& -3.61	& ~ \\	
i 	& 17.92729	& 50.0863	& -4.58	& ~ \\	
\hline
\end{tabular}
\end{table}

\onllongtab{3}{
\begin{longtable}{ccccc}
\caption{The absolute positions, LSR velocities, and fluxes of the centroids of all 6.7~GHz methanol maser spots measured using EVN.}\\
\hline  \hline
RA & DEC & v$_\mathrm{LSR}$ & I$_\mathrm{peak}$ & F\\
(J2000) & (J2000) & [km\,s$^{-1}$] & [Jy\,beam$^{-1}$] & [Jy]\\
\hline
\endfirsthead
\caption{Continued.} \\
\hline
RA & DEC & v$_\mathrm{LSR}$ & I$_\mathrm{peak}$ & F\\
(J2000) & (J2000) & [km\,s$^{-1}$] & [Jy\,beam$^{-1}$] & [Jy]\\
\hline
\endhead
\hline
\endfoot
\hline
\endlastfoot
22:56:17.91048 & 62:01:49.5551 & -1.51 & 0.5 & 0.9 \\
22:56:17.91041 & 62:01:49.5552 & -1.60 & 3.2 & 5.9 \\
22:56:17.91032 & 62:01:49.5553 & -1.68 & 6.6 & 12.8 \\
22:56:17.91022 & 62:01:49.5552 & -1.77 & 5.9 & 12.8 \\
22:56:17.91009 & 62:01:49.5551 & -1.86 & 2.6 & 6.4 \\
22:56:17.96341 & 62:01:49.4451 & -1.95 & 0.3 & 0.4 \\
22:56:17.90334 & 62:01:49.5723 & -1.95 & 0.6 & 1.3 \\
22:56:17.91007 & 62:01:49.5546 & -1.95 & 0.5 & 0.9 \\
22:56:17.90439 & 62:01:49.5862 & -2.03 & 2.6 & 4.1 \\
22:56:17.96453 & 62:01:49.4402 & -2.03 & 0.4 & 0.5 \\
22:56:17.90438 & 62:01:49.5862 & -2.03 & 2.6 & 4.1 \\
22:56:17.96457 & 62:01:49.4401 & -2.12 & 0.8 & 0.8 \\
22:56:17.90441 & 62:01:49.5867 & -2.12 & 8.7 & 14.7 \\
22:56:17.96460 & 62:01:49.4400 & -2.21 & 1.2 & 1.4 \\
22:56:17.90447 & 62:01:49.5867 & -2.21 & 16.1 & 26.4 \\
22:56:17.96462 & 62:01:49.4398 & -2.30 & 1.9 & 2.2 \\
22:56:17.90457 & 62:01:49.5865 & -2.30 & 19.5 & 31.3 \\
22:56:17.96464 & 62:01:49.4397 & -2.39 & 2.0 & 2.1 \\
22:56:17.90485 & 62:01:49.5854 & -2.39 & 20.4 & 35.4 \\
22:56:17.96210 & 62:01:49.4376 & -2.47 & 1.7 & 2.3 \\
22:56:17.89923 & 62:01:49.6113 & -2.47 & 1.4 & 5.8 \\
22:56:17.90515 & 62:01:49.5843 & -2.47 & 28.9 & 44.0 \\
22:56:17.96204 & 62:01:49.4369 & -2.56 & 2.5 & 3.8 \\
22:56:17.89716 & 62:01:49.6178 & -2.56 & 2.5 & 5.7 \\
22:56:17.89870 & 62:01:49.6124 & -2.56 & 3.8 & 9.8 \\
22:56:17.90530 & 62:01:49.5840 & -2.56 & 27.6 & 39.4 \\
22:56:17.96174 & 62:01:49.4369 & -2.65 & 1.3 & 4.7 \\
22:56:17.90541 & 62:01:49.5839 & -2.65 & 13.3 & 18.6 \\
22:56:17.89862 & 62:01:49.6124 & -2.65 & 3.0 & 6.3 \\
22:56:17.89690 & 62:01:49.6186 & -2.65 & 3.4 & 12.0 \\
22:56:17.96084 & 62:01:49.4379 & -2.74 & 2.2 & 2.8 \\
22:56:17.90552 & 62:01:49.5839 & -2.74 & 3.2 & 4.5 \\
22:56:17.89637 & 62:01:49.6200 & -2.74 & 2.4 & 10.7 \\
22:56:17.96348 & 62:01:49.4266 & -2.82 & 0.8 & 2.4 \\
22:56:17.96083 & 62:01:49.4371 & -2.82 & 2.0 & 2.6 \\
22:56:17.90566 & 62:01:49.5838 & -2.82 & 0.6 & 0.9 \\
22:56:17.89593 & 62:01:49.6215 & -2.82 & 1.1 & 4.8 \\
22:56:17.89533 & 62:01:49.6224 & -2.82 & 1.4 & 3.7 \\
22:56:17.96363 & 62:01:49.4259 & -2.91 & 1.1 & 2.5 \\
22:56:17.96082 & 62:01:49.4364 & -2.91 & 0.9 & 1.1 \\
22:56:17.91577 & 62:01:49.5567 & -2.91 & 0.2 & 0.4 \\
22:56:17.89509 & 62:01:49.6233 & -2.91 & 0.5 & 1.3 \\
22:56:17.96364 & 62:01:49.4260 & -3.00 & 0.8 & 1.4 \\
22:56:17.91569 & 62:01:49.5581 & -3.00 & 0.7 & 1.2 \\
22:56:17.96362 & 62:01:49.4263 & -3.09 & 0.3 & 0.4 \\
22:56:17.91564 & 62:01:49.5588 & -3.09 & 0.8 & 1.3 \\
22:56:17.96988 & 62:01:49.4386 & -3.18 & 0.2 & 0.2 \\
22:56:17.91560 & 62:01:49.5592 & -3.18 & 0.4 & 0.6 \\
22:56:17.96988 & 62:01:49.4387 & -3.26 & 0.3 & 0.4 \\
22:56:17.87414 & 62:01:49.7124 & -3.26 & 0.2 & 1.2 \\
22:56:17.96987 & 62:01:49.4387 & -3.35 & 0.3 & 0.3 \\
22:56:17.87506 & 62:01:49.7049 & -3.35 & 0.4 & 2.1 \\
22:56:17.87396 & 62:01:49.7105 & -3.35 & 0.6 & 2.0 \\
22:56:17.96982 & 62:01:49.4387 & -3.44 & 0.2 & 0.2 \\
22:56:17.87555 & 62:01:49.7023 & -3.44 & 0.7 & 1.4 \\
22:56:17.87421 & 62:01:49.7063 & -3.44 & 0.9 & 2.4 \\
22:56:17.87248 & 62:01:49.7294 & -3.44 & 0.2 & 0.2 \\
22:56:17.87433 & 62:01:49.7038 & -3.53 & 0.8 & 1.3 \\
22:56:17.86686 & 62:01:49.7460 & -3.53 & 0.8 & 1.5 \\
22:56:17.86047 & 62:01:49.8186 & -3.53 & 1.2 & 1.8 \\
22:56:17.87135 & 62:01:49.9011 & -3.53 & 0.2 & 0.3 \\
22:56:17.86688 & 62:01:49.7449 & -3.61 & 2.8 & 6.2 \\
22:56:17.86038 & 62:01:49.8191 & -3.61 & 2.5 & 3.6 \\
22:56:17.86265 & 62:01:49.8313 & -3.61 & 0.5 & 0.8 \\
22:56:17.87133 & 62:01:49.9007 & -3.61 & 0.4 & 0.8 \\
22:56:17.87820 & 62:01:49.9047 & -3.61 & 0.2 & 0.4 \\
22:56:17.86687 & 62:01:49.7442 & -3.70 & 3.8 & 9.3 \\
22:56:17.86026 & 62:01:49.8200 & -3.70 & 1.6 & 2.1 \\
22:56:18.09933 & 62:01:49.4214 & -3.79 & 0.7 & 1.6 \\
22:56:17.86586 & 62:01:49.7630 & -3.79 & 0.7 & 2.0 \\
22:56:17.86687 & 62:01:49.7431 & -3.79 & 2.4 & 5.8 \\
22:56:18.09493 & 62:01:49.3971 & -3.88 & 0.9 & 3.3 \\
22:56:18.09913 & 62:01:49.4211 & -3.88 & 0.8 & 1.6 \\
22:56:17.86692 & 62:01:49.7423 & -3.88 & 1.1 & 2.9 \\
22:56:18.09559 & 62:01:49.4004 & -3.97 & 3.1 & 9.8 \\
22:56:17.86710 & 62:01:49.7417 & -3.97 & 0.3 & 0.7 \\
22:56:17.85946 & 62:01:49.7724 & -3.97 & 0.3 & 0.9 \\
22:56:18.09627 & 62:01:49.4033 & -4.05 & 5.6 & 21.7 \\
22:56:18.09648 & 62:01:49.4049 & -4.14 & 9.5 & 20.2 \\
22:56:18.09652 & 62:01:49.4054 & -4.23 & 8.3 & 12.3 \\
22:56:17.92613 & 62:01:50.2357 & -4.32 & 0.2 & 1.5 \\
22:56:18.09651 & 62:01:49.4055 & -4.32 & 2.7 & 3.5 \\
22:56:17.92598 & 62:01:50.2361 & -4.40 & 0.3 & 1.7 \\
22:56:17.92753 & 62:01:50.0846 & -4.49 & 0.4 & 1.4 \\
22:56:17.92729 & 62:01:50.0863 & -4.58 & 0.6 & 2.1 \\
22:56:17.92688 & 62:01:50.0894 & -4.67 & 0.3 & 0.7 \\
22:56:17.92290 & 62:01:50.1160 & -4.76 & 0.2 & 0.8 \\
\hline
\end{longtable}
}

\subsection{12.2~GHz results}\label{12ghzresults}
The 12.2~GHz methanol maser emission is grouped in two distinct clumps detected in all four epochs, a spectra of one epoch is shown in Fig. \ref{maserspectra}. The two 12.2~GHz maser clumps are associated with the 6.7~GHz maser clumps \emph{a} and \emph{c} in the 6.7~GHz maser map (Fig. \ref{cepamap}) and their absolute positions are reported in Table \ref{12ghzmasertable}. The first two of our epochs are separated by only 20~days to those of Mos09 and for these our absolute positions agree with only a small $\sim$1.6~mas offset in declination. This offset is probably due to different calibration strategies, in particular we do not include the rigorous tropospheric corrections of Mos09. The absolute astrometry of Min01 (30~mas accuracy) is not quite good enough for a comparison on the mas level, and although their 12.2~GHz maser feature seems to be associated with a different 6.7~GHz feature (\emph{d}), the general morphology and velocity field look very similar.

\begin{figure}
\centering
\includegraphics[width=9cm]{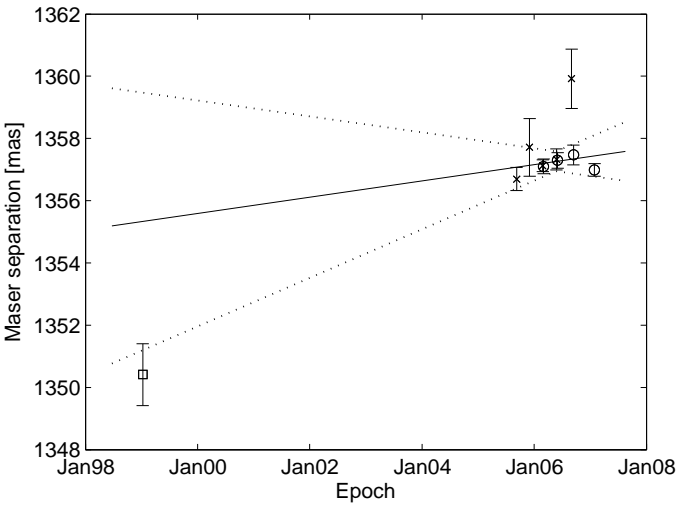}
\caption{Separation of the 12.2~GHz maser clumps in mas. Our four epochs are marked as circles, also shown are previous measurements by Min01 (square) and Mos09 (crosses). The solid line indicates our best fit and the dotted lines the errors of the fit.}
\label{masersep}
\end{figure}

The separation of the 12.2~GHz maser clumps for the four epochs of our observations, the five epochs of Mos09, and the epoch of Min01 is shown in Fig. \ref{masersep}. The measured separation of the two clumps ($\sim$1.357\arcsec) agrees well with those of Mos09. The measurement by Min01 shows a separation of $\sim$6~mas less, though due to the large separation in time we cannot be certain that it is the same parcel of gas. By fitting a straight line to the data points (with the exception of the Min01 data point which is only shown for completeness) we find the separation to be increasing by $0.87 \pm 1.72$~\kms\ ($0.26 \pm 0.52$~mas\,year$^{-1}$). Mos09 found the separation to be increasing by 6.3~\kms, and by looking at Fig. \ref{masersep}, it is clear that this value is largely determined by the separation derived from his last epoch which is inconsistent with our measurements. In conclusion, taking into consideration the error bars of our measurements, we do not see any significant change in the separation over the time period ($\sim$16.5 months) spanned by Mos09 and our observations.

Although taken at $\sim$1.5~year separated epochs, the 12.2~GHz masers are closely ($20 - 40$~mas) associated with the 6.7~GHz methanol masers and show the same velocity field. However, if the emission was arising from the same gas, the observed offset would imply velocities between 50 and 100~\kms. Such high velocities should result in measurable proper motions over the time span covered by our observations and we therefore believe that the separation is due to either variability, or more likely, that the 6.7 and 12.2~GHz maser emission do not exactly arise in the same gas.

We compared whether there was any association between the methanol masers and any other masing species (water or hydroxyl) in Cep A. Only in one position, namely in the elongated structure to the East did we find a close association with water masers. The methanol maser emission is spatially coincident $<5$~mas to the 22~GHz water masers observed by \citet{vlemmings06, torrelles10a}. The 6.7 and 12.2~GHz methanol masers are centred at -4.2~\kms, in contrast, the 22~GHz water masers are observed at $\sim-$14~\kms. This velocity offset will be further discussed in Sec. \ref{discussion}.

\begin{table}[h]
\caption{The absolute positions and LSR velocities of 12.2~GHz methanol maser clumps measured using VLBA for the four different epochs (I-IV).}
\label{12ghzmasertable}
\centering
\begin{tabular}{lccc}
\hline \hline
Epoch & RA &  Dec & Velocity \\
~ &(J2000) & (J2000) & [{km\,s$^{-1}$}]    \\
~ & 22:56: & 62:01:   & ~ \\
\hline
   I 	& 18.097074 		& 49.39635 		& -4.1	\\
	& 17.905360 		& 49.54696 		& -1.8	\\
   II	& 18.097174 		& 49.39681 		& -4.1	\\
	& 17.905432 		& 49.54742 		& -1.8	\\
   III 	& 18.096963 		& 49.39687 		& -4.1	\\
	& 17.905200 		& 49.54781 		& -1.8	\\
   IV	& 18.096775 		& 49.39301 		& -4.1	\\
	& 17.905086 		& 49.54415 		& -1.8	\\
\hline
\end{tabular}
\end{table}

\section{Analysis}\label{analysis}
\subsection{Ring model}
The methanol maser emission seems to arise in a large-scale, ring-like structure close to or in the equatorial region of the high-mass object driving the HW2 jet. Such elliptical structures have recently been discovered in a substantial fraction of methanol masers studied in detail with VLBI \citep{bartkiewicz05b, bartkiewicz09}. To model our data we have done a least square fit of an ellipse to the positions of our 6.7~GHz methanol maser spots. We solve for the ellipse's semi-major and semi-minor axes, the position angle and for the position offset with respect to the 7~mm continuum object identified by \citet{curiel06}. Furthermore, assuming that the maser emission is arising in a circular ring structure, we convert the fitted ellipse to a position angle of the minor axis, a radius and an inclination of the ring. 

\begin{figure*}
\centering
{
    \includegraphics[width=0.45\textwidth]{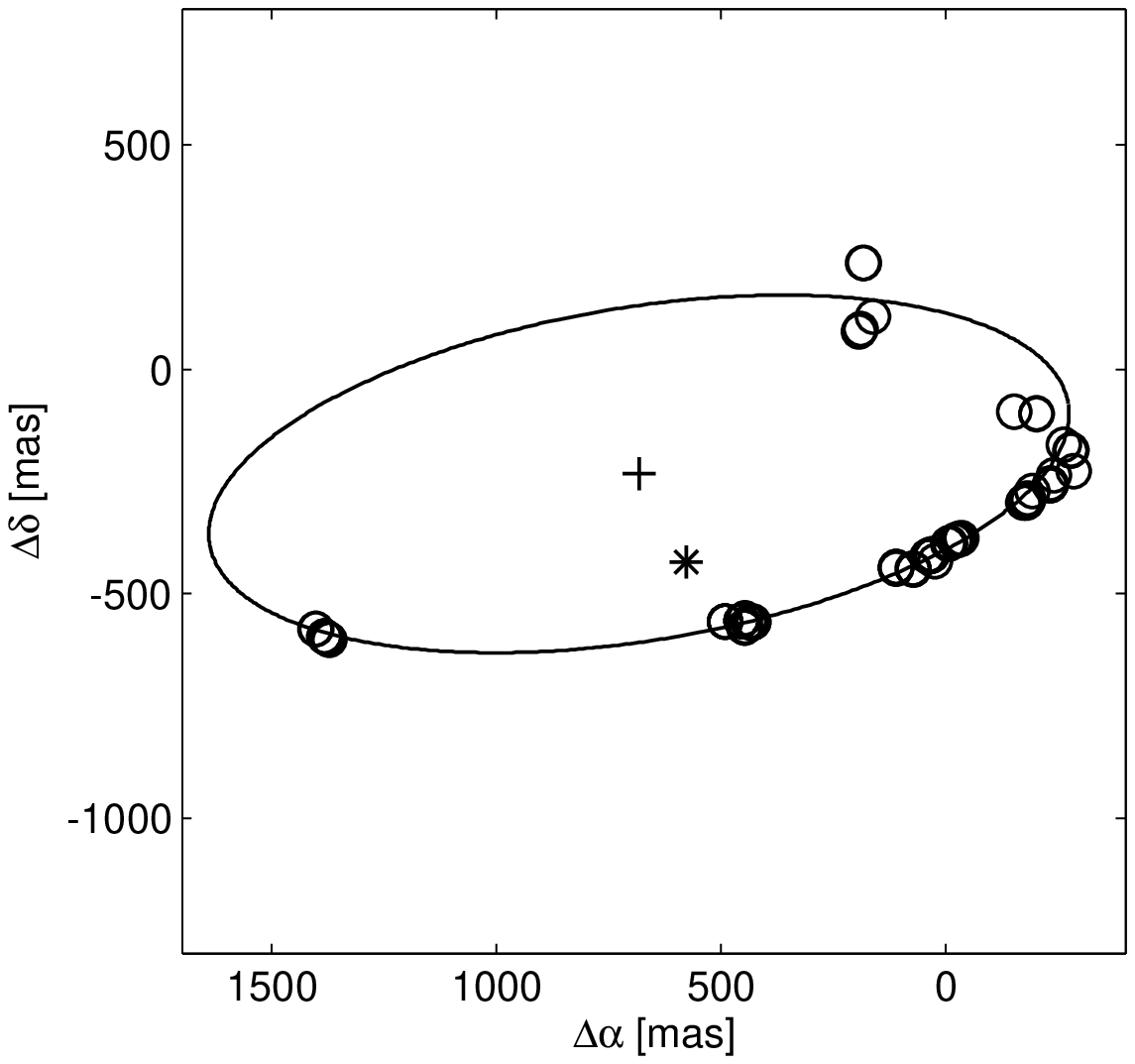}
}
{
    \includegraphics[width=0.43\textwidth]{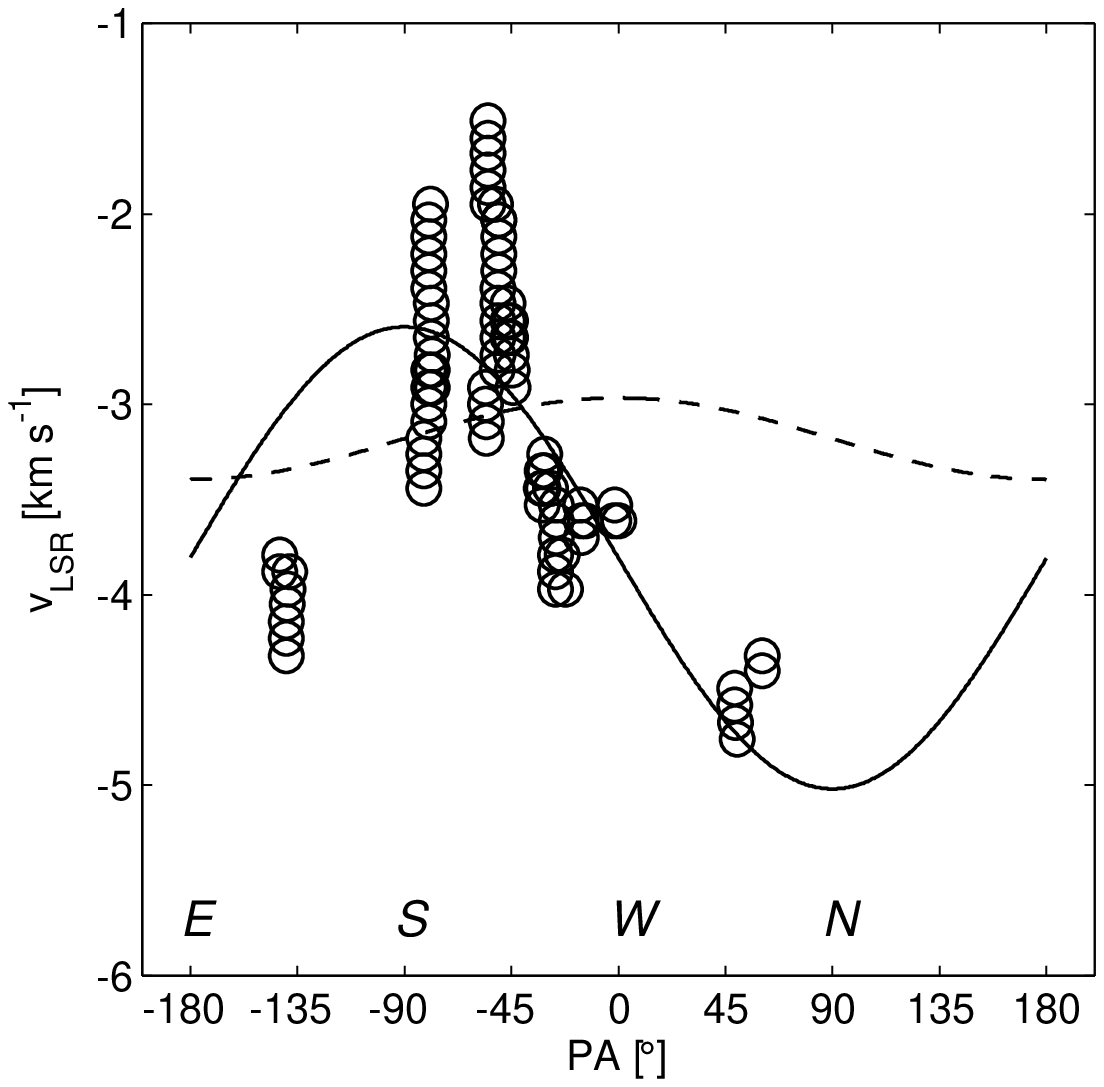}
}
\caption{Fit of the ring model: Left panel shows the maser positions in circles and the fitted ellipse, with the $+$ sign designating the centre of the ellipse. The star indicates the position of the 7~mm continuum object identified by \citet{curiel06}. Right panel shows in the measured velocities (circles) as a function of the position angle (PA) in our ring model. The solid line shows the fit to a radial velocity model and the dashed line shows the result of a Keplerian velocity model.}
\label{ringmodel}
\end{figure*}

The best fit and model results are presented in Fig. \ref{ringmodel}, with offsets relative to the phase centre (Table \ref{obstable}). As one can clearly see the maser spots are consistent with an ellipse only slightly offset from the 7~mm continuum object ($\Delta \alpha=-106$~mas, $\Delta \delta=197$~mas). The semi-major axis of the fitted ellipse is 968~mas (678~AU), and the semi-minor axis 371~mas (260~AU). We find the position angle of the minor axis to be 9.3\degr. The radio jet and the large scale molecular outflow have a position angle of 45\degr\ and given that these are completely different structures we note that they seem to be pointing in the same general direction. Assuming the masers are in an inclined ring we find an inclination of 67.5\degr, which is slightly larger than the best fit of $62$\degr\ that was found for the dust and molecular disk \citep{patel05}. Both the position angle and the inclination of the ring are in good agreement with the MERLIN 6.7~GHz maser observations (i=71\degr, pa=12\degr) by \citet{vlemmings10a}.

The ring model fitted to the methanol maser positions is used to compare two simple models that have different kinematic signatures: radial velocities ($v(PA)=v_\mathrm{sys}+v_\mathrm{rad} \times \mathrm{sin}(i)\ \mathrm{sin}(PA)$) and Keplerian rotation ($v(PA)=v_\mathrm{sys}+v_\mathrm{rot} \times \mathrm{sin}(i)\ \mathrm{cos}(PA)$), Fig. \ref{ringmodel}. For the radial velocity model we find a system velocity of $v_\mathrm{sys}=-3.8$~\kms\ and a radial velocity $v_\mathrm{rad}=-1.3$~\kms. For the rotation model we find a $v_\mathrm{sys}=-3.1$~\kms and a rotation velocity, $v_\mathrm{rot}=0.2$~\kms. Assuming Keplerian rotation this would correspond to a mass of $\sim$0.04~\Msun, which is much too low for any reasonable model of the mass enclosed, we therefore rule out this possibility. Moreover, the rotation model does not fit. In Fig. \ref{cepamap} the maser regions to the East and West show similar velocities and the high velocity is seen towards the centre, this is a sign of radial motions, not of rotation. Whether the radial motions can be attributed to infall or outflow, depends on the sign of the inclination; if the half-ellipse from which we see the maser emission is the front-side, the signature is that of infall. In Sect. \ref{discussion} we argue that most observed effects are indeed consistent with this. With the long baselines of the EVN the maser emission is broken up in many individual spots, it is however clear we see the same structure as observed with the MERLIN \citep{vlemmings10a}.

\subsection{Parallax - distance}
As we have independent measurements of the absolute positions of the 12.2~GHz maser clumps, we can verify the important findings of Mos09 on the distance to Cep~A. A significant difference with their processing setup is that we have not included an estimate of the tropospheric errors \citep{reid09a}, probably resulting in the 1.6~mas offset in declination. Moreover, we have only one of the two calibrators in common. Otherwise the data is of very similar quality and we can fit our observations with the same parameters for proper motion and parallax. As was the case for Mos09 the right ascension data is very inconclusive, but the declination data result in an unambiguous determination of the parallax. Combining the measurements, taking out the aforementioned 1.6~mas offset, we can find a consistent result, $\pi=1.567 \pm 0.087$~mas ($0.64^{+0.04}_{-0.03}$~kpc). However, we think the Mos09 value of $0.70^{+0.04}_{-0.04}$~kpc is probably still preferred because of a more thorough treatment of systematic errors.

\section{Discussion}\label{discussion}
Although individual maser spots are small scale signposts of specific excitation conditions, the morphology and velocity field of the maser emission show that they occur within a single large scale ($\sim1350$~AU) ring structure around the high mass protostar HW2. Such structures have recently been found in a large number of methanol maser sources \citep{bartkiewicz09}, indicating that at least a fraction of the methanol masers are associated with specific conditions at a few hundred AU from the central source. Cep~A HW2 is unique in that we can determine both the inclination of the ring as seen in methanol masers and the outflow direction as seen in radio continuum. This allows us, together with the large scale ($\sim 1$\arcmin) molecular outflow, to constrain the geometry of the system. The molecular outflow to the NE is blueshifted, pointing towards us and the outflow to the SW is redshifted, pointing away from us \citep{gomez99a}. The orientation of the ring and its location suggest that the main part of the methanol maser emission (the large arc) arises on the near side of the jet, and the maser emission in the centre is seen in projection against the south western radio lobe. The maser emission to NNE of the centre of the ring (from the presumed far side) are only just detected with the EVN.

We can consider a few causes for the lesser brightness of the maser emission from the presumed far side of the ring. Assuming optically thin emission at 7~mm, \citet{curiel06} found an electron density of (1--5)$\times10^4$~cm$^{-3}$ in the radio jet, not enough for the emission to be optically thick at 6.7~GHz. For an optical depth of $\tau=1$ at 6.7~GHz an emission measure of $>1.7 \times 10^8$~cm$^{-6}$\,pc$^{-1}$ is required, assuming an electron temperature of $10^4$~K. This corresponds to an electron density $>1.5 \times 10^5$~cm$^{-3}$ within the ring, a few times higher than the value found by \citet{curiel06}. However, the rising spectra from 3.6~cm to 7~mm suggest that the optical depth is significant at 6.7~GHz, and it is therefore reasonable that optical depth effects play some role in at least part of the ring. A second cause for the maser emission from the near side of the ring to be brighter than the far side is the background radiation from the central object enhancing the outward radial maser brightness. This is supported by the lower brightness/larger scale maser emission arising from the far side of the ring and observed with MERLIN \citep{vlemmings10a} which may to a high degree be filtered out on the much longer EVN baselines. For the remainder of the discussion we assume that the maser emission delineating the arc arises on the near side of the jet. 

Then, the velocity field of the masers, with the blueshifted masers towards the sides (tangents) and the more redshifted masers towards the centre indicates an infall signature. No clear signs of rotation are seen. We note that similar velocity ``spurs'' as seen in Fig. \ref{ringmodel} have been seen in other maser sources \citep{bartkiewicz09} and seem inherent to the clumpy maser emission. In fact, the velocity gradient in individual maser features could be the result of gas starting to be entrained in the rotation of the accretion disk. The low velocity maser emission ($\sim -4.6$~\kms) that we just detect and that is seen at shorter baselines falls within the third quarter of the velocity diagram of Fig. \ref{ringmodel} where we expect velocities between -4.2 and -5.9~\kms. Although we cannot completely rule out that the 6.7 and 12.2~GHz methanol maser emission do not arise from within the same parcels of gas, the close association and similar velocities indicate that at least they occur in the same large scale structure in the equatorial region of HW2. The velocity field of the masers suggests a modest infall, although the fact that we do not measure any significant change in the separation between the 12.2~GHz clumps, seems to indicate that they do not move relative to each other. We tentatively identify this ring structure as the interface between the larger scale accretion flow and a circumstellar disk. This scenario is further supported by the large scale magnetic field perpendicular to the ring structure \citep{vlemmings10a}.

Recent observations of the 22~GHz water masers in the region of Cep~A HW2 show that the water masers trace a relatively slow, wide-angle (102\degr) outflow to the NE, present at the same time and on similar spatial scales ($\sim $1\arcsec) as the highly collimated jet seen in radio continuum with outflow velocities in excess of 500~\kms\ \citep{torrelles10a}. In the East the methanol masers are closely associated with these water masers. However, the proper motions and velocity fields of the methanol masers show that they do not trace the same part of the structure. A similar close association between water and methanol masers with a different velocity field has previously been seen in at least one other source (AFGL 5142). The results by \citet{goddi07a} indicate that the water masers trace expansion, possibly related to the disk-wind interface, whereas the methanol masers seem to trace an infall. This could be understood from the different conditions required; models of maser excitation show that water masers requires a denser ($n>10^9$~cm$^{-3}$) and warmer medium in which the methanol masers are quenched \citep{elitzur89a, cragg05a}. In Cep~A, the previously noted velocity offset of $\sim$10~\kms\ between the two maser species suggest a shock front in which the methanol masers occur outside (upstream) of the shock interface, in the infalling gas, and the water masers occur further downstream or behind the shock front, where conditions are more favourable for water maser excitation. 

Monitoring of the 6.7~GHz methanol masers have shown the maser emission to be not only variable but also that the variability is synchronised and anti-correlated between different sections of the ring \citep{sugiyama08b}. As the intensity in the blueshifted masers to the East and West decreases the intensity of the redshifted masers towards the centre increases. The authors ascribe the variability to changes to the excitation conditions in the environment in which the maser emission occur, and that the masers towards the centre are closer to the central object than the blueshifted masers. The variability could also be consistent with a model in which amplification of the HW2 radio continuum is important. In this scenario the radial amplification will be enhanced as the intensity of the central object increases. The increase in radial amplification would enhance the beaming, and consequently decrease the tangential amplification. This explanation could be tested by simultaneously monitoring the maser and the radio continuum emission (with for example the e-VLA or e-MERLIN).

\section{Conclusions}\label{conclusions}
We find the methanol maser emission in Cep~A to arise in a large scale ring structure straddling the waist of the thermal jet HW2. We find the ring to have a radius of 678~AU centred close to the previously identified central object driving the HW2 radio jet. Further, the minor axis of the ring is in the general direction of the radio jet, the large scale molecular outflow, and the large scale magnetic field. The geometry suggests that the ring is inclined with the NE part towards us and the maser emission arises on the near side of the ring. This leads us to interpret the velocity field of the methanol masers as showing signs of modest infall although we do not see any change in the separation of the maser clumps. Based on the geometry, velocity field, and large velocity offset from the water masers we propose that the methanol masers occur in a shock interface of the infalling gas, on the interface with the accretion disk. We also suggest that amplification of background radio continuum is important in the appearance and variability of the methanol maser emission. Our independent fit of the parallax to Cep~A HW2 confirms the previously measured parallax of Mos09. 

\begin{acknowledgements}
This research was supported by the EU Framework 6 Marie Curie Early Stage 
Training programme under contract number MEST-CT-2005-19669 ``ESTRELA''. WV acknowledges support by the {\it Deutsche Forschungsgemeinschaft} through the Emmy Noether Research grant VL 61/3-1.
We are grateful to Luca Moscadelli and collaborators for kindly sharing their results (Mos09) in digital form with us.
\end{acknowledgements}

\bibliographystyle{aa}
\bibliography{kalle}

\begin{thebibliography}{34}
\expandafter\ifx\csname natexlab\endcsname\relax\def\natexlab#1{#1}\fi

\bibitem[{{Bartkiewicz} {et~al.}(2005{\natexlab{a}}){Bartkiewicz}, {Szymczak},
  {Cohen}, \& {Richards}}]{bartkiewicz05a}
{Bartkiewicz}, A., {Szymczak}, M., {Cohen}, R.~J., \& {Richards}, A.~M.~S.
  2005{\natexlab{a}}, \mnras, 361, 623

\bibitem[{{Bartkiewicz} {et~al.}(2005{\natexlab{b}}){Bartkiewicz}, {Szymczak},
  \& {van Langevelde}}]{bartkiewicz05b}
{Bartkiewicz}, A., {Szymczak}, M., \& {van Langevelde}, H.~J.
  2005{\natexlab{b}}, \aap, 442, L61

\bibitem[{{Bartkiewicz} {et~al.}(2009){Bartkiewicz}, {Szymczak}, {van
  Langevelde}, {Richards}, \& {Pihlstr{\"o}m}}]{bartkiewicz09}
{Bartkiewicz}, A., {Szymczak}, M., {van Langevelde}, H.~J., {Richards},
  A.~M.~S., \& {Pihlstr{\"o}m}, Y.~M. 2009, \aap, 502, 155

\bibitem[{{Bourke} {et~al.}(2006){Bourke}, {van Langevelde}, {Harvey-Smith}, \&
  {Golden}}]{bourke06a}
{Bourke}, S., {van Langevelde}, H.~J., {Harvey-Smith}, L., \& {Golden}, A.
  2006, in Proceedings of the 8th European VLBI Network Symposium

\bibitem[{{Cragg} {et~al.}(2005){Cragg}, {Sobolev}, \& {Godfrey}}]{cragg05a}
{Cragg}, D.~M., {Sobolev}, A.~M., \& {Godfrey}, P.~D. 2005, \mnras, 360, 533

\bibitem[{{Curiel} {et~al.}(2006){Curiel}, {Ho}, {Patel}, {Torrelles},
  {Rodr{\'{\i}}guez}, {Trinidad}, {Cant{\'o}}, {Hern{\'a}ndez}, {G{\'o}mez},
  {Garay}, \& {Anglada}}]{curiel06}
{Curiel}, S., {Ho}, P.~T.~P., {Patel}, N.~A., {et~al.} 2006, \apj, 638, 878

\bibitem[{{Elitzur} {et~al.}(1989){Elitzur}, {Hollenbach}, \&
  {McKee}}]{elitzur89a}
{Elitzur}, M., {Hollenbach}, D.~J., \& {McKee}, C.~F. 1989, \apj, 346, 983

\bibitem[{{Evans} {et~al.}(1981){Evans}, {Slovak}, {Becklin}, {Beichman},
  {Gatley}, {Werner}, {Hildebrand}, {Keene}, \& {Whitcomb}}]{evans81}
{Evans}, II, N.~J., {Slovak}, M.~H., {Becklin}, E.~E., {et~al.} 1981, \apj,
  244, 115

\bibitem[{{Goddi} {et~al.}(2007){Goddi}, {Moscadelli}, {Sanna}, {Cesaroni}, \&
  {Minier}}]{goddi07a}
{Goddi}, C., {Moscadelli}, L., {Sanna}, A., {Cesaroni}, R., \& {Minier}, V.
  2007, \aap, 461, 1027

\bibitem[{{G{\'o}mez} {et~al.}(1999){G{\'o}mez}, {Sargent}, {Torrelles}, {Ho},
  {Rodr{\'{\i}}guez}, {Cant{\'o}}, \& {Garay}}]{gomez99a}
{G{\'o}mez}, J.~F., {Sargent}, A.~I., {Torrelles}, J.~M., {et~al.} 1999, \apj,
  514, 287

\bibitem[{{Hill} {et~al.}(2005){Hill}, {Burton}, {Minier}, {Thompson}, {Walsh},
  {Hunt-Cunningham}, \& {Garay}}]{hill05}
{Hill}, T., {Burton}, M.~G., {Minier}, V., {et~al.} 2005, \mnras, 363, 405

\bibitem[{{Hughes} \& {Wouterloot}(1984)}]{hw84}
{Hughes}, V.~A. \& {Wouterloot}, J.~G.~A. 1984, \apj, 276, 204

\bibitem[{{Jim{\'e}nez-Serra} {et~al.}(2009){Jim{\'e}nez-Serra},
  {Mart{\'{\i}}n-Pintado}, {Caselli}, {Mart{\'{\i}}n},
  {Rodr{\'{\i}}guez-Franco}, {Chandler}, \& {Winters}}]{jimenez-serra09a}
{Jim{\'e}nez-Serra}, I., {Mart{\'{\i}}n-Pintado}, J., {Caselli}, P., {et~al.}
  2009, \apjl, 703, L157

\bibitem[{{Kettenis} {et~al.}(2006){Kettenis}, {van Langevelde}, {Reynolds}, \&
  {Cotton}}]{kettenis06a}
{Kettenis}, M., {van Langevelde}, H.~J., {Reynolds}, C., \& {Cotton}, B. 2006,
  in Astronomical Society of the Pacific Conference Series, Vol. 351,
  Astronomical Data Analysis Software and Systems XV, ed. {C.~Gabriel,
  C.~Arviset, D.~Ponz, \& S.~Enrique}, 497--+

\bibitem[{{McKee} \& {Tan}(2003)}]{mckee03a}
{McKee}, C.~F. \& {Tan}, J.~C. 2003, \apj, 585, 850

\bibitem[{{Minier} {et~al.}(2002){Minier}, {Booth}, \& {Conway}}]{minier02}
{Minier}, V., {Booth}, R.~S., \& {Conway}, J.~E. 2002, \aap, 383, 614

\bibitem[{{Minier} {et~al.}(2001){Minier}, {Conway}, \& {Booth}}]{minier01}
{Minier}, V., {Conway}, J.~E., \& {Booth}, R.~S. 2001, \aap, 369, 278

\bibitem[{{Moscadelli} {et~al.}(2009){Moscadelli}, {Reid}, {Menten},
  {Brunthaler}, {Zheng}, \& {Xu}}]{moscadelli09}
{Moscadelli}, L., {Reid}, M.~J., {Menten}, K.~M., {et~al.} 2009, \apj, 693, 406

\bibitem[{{Norris} {et~al.}(1998){Norris}, {Byleveld}, {Diamond}, {Ellingsen},
  {Ferris}, {Gough}, {Kesteven}, {McCulloch}, {Phillips}, {Reynolds},
  {Tzioumis}, {Takahashi}, {Troup}, \& {Wellington}}]{norris98a}
{Norris}, R.~P., {Byleveld}, S.~E., {Diamond}, P.~J., {et~al.} 1998, \apj, 508,
  275

\bibitem[{{Patel} {et~al.}(2005){Patel}, {Curiel}, {Sridharan}, {Zhang},
  {Hunter}, {Ho}, {Torrelles}, {Moran}, {G{\'o}mez}, \& {Anglada}}]{patel05}
{Patel}, N.~A., {Curiel}, S., {Sridharan}, T.~K., {et~al.} 2005, \nat, 437, 109

\bibitem[{{Pestalozzi} {et~al.}(2004){Pestalozzi}, {Elitzur}, {Conway}, \&
  {Booth}}]{pestalozzi04}
{Pestalozzi}, M.~R., {Elitzur}, M., {Conway}, J.~E., \& {Booth}, R.~S. 2004,
  \apjl, 603, L113

\bibitem[{{Reid} {et~al.}(2009){Reid}, {Menten}, {Brunthaler}, {Zheng},
  {Moscadelli}, \& {Xu}}]{reid09a}
{Reid}, M.~J., {Menten}, K.~M., {Brunthaler}, A., {et~al.} 2009, \apj, 693, 397

\bibitem[{{Rygl} {et~al.}(2010){Rygl}, {Brunthaler}, {Reid}, {Menten}, {van
  Langevelde}, \& {Xu}}]{rygl10a}
{Rygl}, K.~L.~J., {Brunthaler}, A., {Reid}, M.~J., {et~al.} 2010, \aap, 511, A2

\bibitem[{{Sugiyama} {et~al.}(2008{\natexlab{a}}){Sugiyama}, {Fujisawa}, {Doi},
  {Honma}, {Isono}, {Kobayashi}, {Mochizuki}, \& {Murata}}]{sugiyama08b}
{Sugiyama}, K., {Fujisawa}, K., {Doi}, A., {et~al.} 2008{\natexlab{a}}, \pasj,
  60, 1001

\bibitem[{{Sugiyama} {et~al.}(2008{\natexlab{b}}){Sugiyama}, {Fujisawa}, {Doi},
  {Honma}, {Kobayashi}, {Bushimata}, {Mochizuki}, \& {Murata}}]{sugiyama08a}
{Sugiyama}, K., {Fujisawa}, K., {Doi}, A., {et~al.} 2008{\natexlab{b}}, \pasj,
  60, 23

\bibitem[{{Sugiyama} {et~al.}(2007){Sugiyama}, {Fujisawa}, {Honma}, {Doi},
  {Mochizuki}, {Murata}, \& {Isono}}]{sugiyama07a}
{Sugiyama}, K., {Fujisawa}, K., {Honma}, M., {et~al.} 2007, in IAU Symposium,
  Vol. 242, IAU Symposium, ed. {J.~M.~Chapman \& W.~A.~Baan}, 176--177

\bibitem[{{Torrelles} {et~al.}(1998){Torrelles}, {G{\'o}mez}, {Garay},
  {Rodr{\'{\i}}guez}, {Curiel}, {Cohen}, \& {Ho}}]{torrelles98}
{Torrelles}, J.~M., {G{\'o}mez}, J.~F., {Garay}, G., {et~al.} 1998, \apj, 509,
  262

\bibitem[{{Torrelles} {et~al.}(2010){Torrelles}, {Patel}, {Curiel},
  {Estalella}, {Gómez}, {Rodriguez}, {Canto}, {Anglada}, {Vlemmings}, {Garay},
  {Raga}, \& {Ho}}]{torrelles10a}
{Torrelles}, J.~M., {Patel}, N., {Curiel}, S., {et~al.} 2010, \mnras, submitted

\bibitem[{{Torrelles} {et~al.}(2007){Torrelles}, {Patel}, {Curiel}, {Ho},
  {Garay}, \& {Rodr{\'{\i}}guez}}]{torrelles07}
{Torrelles}, J.~M., {Patel}, N.~A., {Curiel}, S., {et~al.} 2007, \apjl, 666,
  L37

\bibitem[{{Vlemmings}(2008)}]{vlemmings08a}
{Vlemmings}, W.~H.~T. 2008, \aap, 484, 773

\bibitem[{{Vlemmings} {et~al.}(2006){Vlemmings}, {Diamond}, {van Langevelde},
  \& {Torrelles}}]{vlemmings06}
{Vlemmings}, W.~H.~T., {Diamond}, P.~J., {van Langevelde}, H.~J., \&
  {Torrelles}, J.~M. 2006, \aap, 448, 597

\bibitem[{{Vlemmings} {et~al.}(2010){Vlemmings}, {Surcis}, {Torstensson}, \&
  {van Langevelde}}]{vlemmings10a}
{Vlemmings}, W.~H.~T., {Surcis}, G., {Torstensson}, K.~J.~E., \& {van
  Langevelde}, H.~J. 2010, \mnras, 404, 134

\bibitem[{{Walsh} {et~al.}(2001){Walsh}, {Bertoldi}, {Burton}, \&
  {Nikola}}]{walsh01a}
{Walsh}, A.~J., {Bertoldi}, F., {Burton}, M.~G., \& {Nikola}, T. 2001, \mnras,
  326, 36

\bibitem[{{Walsh} {et~al.}(1998){Walsh}, {Burton}, {Hyland}, \&
  {Robinson}}]{walsh98}
{Walsh}, A.~J., {Burton}, M.~G., {Hyland}, A.~R., \& {Robinson}, G. 1998,
  \mnras, 301, 640

\end{thebibliography}
\end{document}